\documentclass[amsmath,amssymb,groupedaddress,floats,aps,prl,twocolumn]{revtex4-1}
\usepackage[latin1]{inputenc}
\usepackage{graphicx}
\usepackage{color}
\usepackage[usenames,dvipsnames]{xcolor}
\makeatletter
\allowdisplaybreaks

\begin{document}
\renewcommand{\floatpagefraction}{1.0}
\newcommand{\gbb}{}
\definecolor{MyGreen}{rgb}{0.0,0.5,0.0}
\renewcommand{\vec}[1]{\mathbf{#1}}

\title{Velocity shear, turbulent saturation, and steep plasma gradients in the scrape-off layer of inner-wall limited tokamaks}
\author{Federico D.~Halpern}\email{federico.halpern@gmail.com}
\author{Paolo Ricci}
\affiliation{\'{E}cole Polytechnique F\'{e}d\'{e}rale de Lausanne (EPFL), Swiss Plasma Center, CH-1015 Lausanne, Switzerland}

\begin{abstract}
The narrow power decay-length ($\lambda_q$), recently found in the scrape-off layer (SOL) of inner-wall limited (IWL) discharges in tokamaks, is studied using 3D, flux-driven, global two-fluid turbulence simulations. The formation of the steep plasma profiles measured is found to arise due to radially sheared $\vec{E}\times\vec{B}$ poloidal flows. A complex interaction between sheared flows and outflowing plasma currents regulates the turbulent saturation, determining the transport levels. We quantify the effects of sheared flows, obtaining theoretical estimates in agreement with our non-linear simulations. Analytical calculations suggest that the IWL $\lambda_q$ is roughly equal to the turbulent correlation length.

%Flux-driven 3D turbulence simulations of the scrape-off layer (SOL) region of inner-wall limited tokamak plasmas show a two-exponential power decay-length ($\lambda_q$) structure in the plasma profiles. It is found that the cross-field density and heat transport in the near-SOL, \emph{i.e.} just outside the confined plasma region, are strongly affected by radially sheared $\vec{E}\times\vec{B}$ poloidal flows. This leads to the formation of the so-called "narrow heat-flux feature" recently observed in Alcator C-Mod, DIII-D, TCV, and other tokamaks. Simulation results and simple analytical estimates support a narrow heat-flux feature $\lambda_q$ nearly independent of the plasma major radius $R$, and strongly decreasing with the poloidal magnetic field.
\end{abstract}

\maketitle
\makeatother
%\section{Introduction}
Sheared flows can significantly affect the properties of turbulence in magnetically confined plasmas. These effects are observed in many plasma configurations, an archetype of such phenomena being the spontaneous formation of the high-confinement (H-)mode barrier at the edge of tokamak plasmas~\cite{Wagner1982}. Turbulent suppression typically occurs when the radial shearing rate of the $\vec{E}\times\vec{B}$ plasma flows, $\omega_{\vec{E}\times\vec{B}}=d v_{\vec{E}\times\vec{B}} /dr$ ($v_{\vec{E}\times\vec{B}}=\vec{E}\times\vec{B}/B^2$), is of the order of the linear growth rate of the turbulent modes~\cite{Bigliari1990, Burrell1997}. Understanding the effects of sheared flows is paramount for attaining a fusion reactor, in particular due to their typically beneficial effects upon plasma energy confinement and stability.

The present letter deals with radially sheared $\vec{E}\times\vec{B}$ flows in the open magnetic field line region of tokamak devices, known as the scrape-off layer (SOL). In this region of the device, the balance between cross-field heat transport against parallel streaming along magnetic field lines gives rise to exponentially decaying power profiles with a characteristic length $\lambda_q=-(d_x\ln q_\|)^{-1}$ ($q_\|\sim nc_s T$, with $c_s=\sqrt{(T_e+T_i)/m_i}$, is the power flowing along the magnetic field lines towards the device walls). As opposed to the confined plasma region, where we seek to use sheared flows to minimize turbulent transport, SOL turbulence can help avoid a too narrow power exhaust channel.

We concentrate on the inner-wall limited (IWL) geometry, where the plasma makes contact with the inner-wall of the device. This configuration will be used as a start-up plasma scenario in ITER before standard X-point configuration is attained~\cite{Pitts2011}. It was originally assumed that the ITER IWL SOL could be described with a single-exponential $\lambda_q$ of a few cm's~\cite{Loarte2007}. Recent IWL experiments demonstrated that the SOL plasma profiles have a double-exponential decay length structure. In effect, in the near-SOL just outside the confined plasma region, $\lambda_q$ is an order of magnitude smaller than expected~\cite{Arnoux2013,Horacek2014,Nespoli2014,Dejarnac2015,Stangeby2015}. We refer to this steep gradient region as the "narrow heat-flux feature". A multi-device study projects that the ITER IWL near-SOL $\lambda_q$ will be about 4mm, and prompted a redesign of the inner-wall tiles to accommodate for the significantly smaller than expected $\lambda_q$~\cite{Kocan2015}. %We point out, in particular, the importance of understanding the device size scaling of $\lambda_q$ for tokamaks, as both the IWL and Eich's H-mode scaling projects $\lambda_q\sim R^0$~\cite{Eich2011}, a formidable challenge for future, reactor class devices.

%It has been recently found that the narrow feature $\lambda_q$ appears to follow the heuristic drift (HD) scaling $\lambda_q\approx (a/R) \rho_{s\theta}$ ($a$ and $R$ are the device minor and major radii, $\rho_{s\theta}= c_s / (eB_\theta/m_i)$ is the poloidal gyroradius)~\cite{Goldston2015}. This is a rather surprising result, as the HD model relies on collisional transport processes, which is counterintuitive with respect to experimental observations of large turbulent fluctuations at the edge of IWL plasmas. On the other hand, the HD model was originally derived to explain $\lambda_q$ in H-mode plasmas with an X-point configuration~\cite{Eich2011}. Thus, the similarity between the $\lambda_q$s in these two geometries may signal towards general physical mechanisms at play in all divertor configurations. 
 
%%%, as well as in dedicated IWL discharges in the Alcator C-Mod tokamak
Herein we demonstrate that the steep gradients in the narrow heat-flux feature can arise due to radially sheared $\vec{E}\times\vec{B}$ poloidal flows present at the interface between the confined plasma region and the SOL. We observe this phenomenon in 3D flux-driven turbulence simulations of plasma dynamics in the IWL configuration. Despite the strongly sheared flows, we find a relative fluctuation amplitude of about $20\%$ within the narrow feature in the simulations, which is consistent with experimental observations. The most peculiar and surprising aspect of the simulated dynamics is the role of sheath currents and their interaction with the sheared turbulent flows in regulating cross-field turbulent transport. Considering these phenomena, we develop a reduced transport model capturing the physical mechanisms at play within the narrow feature. The resulting $\lambda_q$ is intimately linked to the turbulent correlation length.

%, and in turn leads to a $\lambda_q$ strongly dependent on the poloidal magnetic field, and nearly independent of the device size. % -- which, in fact, resembles the Eich/Goldston scalings~\cite{Eich2011,Goldston2015}. 

%\section{Model}\label{sec_model}

The formation of a narrow heat-flux feature is demonstrated using 3D flux-driven turbulence simulations of plasma dynamics in the IWL configuration. The non-linear simulations allow us to extract and understand the variation of the near-SOL $\lambda_q$ with the plasma parameters. We make use of the drift-reduced Braginskii equations~\cite{Zeiler1997}, which arise from applying the orderings $d/dt\ll\omega_{ci}$ ($\omega_{ci}=eB/m_i$ is the ion gyrofrequency) and $k_\bot\gg k_\|$ to the Braginskii fluid equations~\cite{Braginskii1965}. We consider the simplest possible model that can be used to recover the narrow heat-flux feature, \emph{i.e.} cold ions, a large aspect ratio torus with circular geometry, and we use the Boussinesq approximation. This entails the physics of drift and ballooning modes, which can be destabilized either by finite resistivity or inertia. The model equations for conservation of density $n$, vorticity $\Omega=\nabla_\bot^2\phi$, parallel electron and and ion velocities $v_{\|e,i}$, and electron temperature $T_e$ read
\begin{align}
%
% Electron density (metric)
%
\frac{dn}{dt} =& \frac{2}{eB}\left[\hat{C}(p_e)-en\hat{C}(\phi)\right]-\nabla\cdot\left(nv_{\|e}\hat{\mathbf{b}} \right)+D_n \nabla^2_\perp n + S_n\label{eq_dens}\\
%
% Vorticity including non-Boussinesq terms (metric)
%
\frac{d\Omega }{dt}=& \frac{2B}{nm_i} \hat{C}(p_e) + \frac{B^2}{nm_i} \nabla \cdot \left( j_\| \hat{\mathbf{b}}\right) - v_{\|i}\nabla_\|\Omega \nonumber\\
& + D_\Omega \nabla^2_\perp \Omega +\frac{B}{3n m_i}\hat{C}(G_i)\label{eq_vort}\\
%
%
% Ohm's law including electromagnetic effects (metric)
%
\frac{dv_{\|e}}{dt} =& \frac{ej_\|}{\sigma_\|m_e} + \frac{e\nabla_{\|}\phi}{m_e}-\frac{\nabla_\| p_e}{nm_e}-\frac{0.71\nabla_{\|}T_e}{m_e}\nonumber\\
&- v_{\| e} \nabla_{\|} v_{\|e}+D_{v_{\|e}}\nabla_\perp^2 v_{\|e} -\frac{2\nabla_{\|}G_e}{3nm_e}\\
%
%
%
% Ion parallel velocity equation (in fact momentum neglecting m_e / m_i terms) (metric)
%
 \frac{dv_{\| i}}{dt} =& -\frac{\nabla_\| p_e}{n} -v_{\|i}\nabla_{\|} v_{\| i}+D_{v_{\|i}}\nabla_\perp^2 v_{\|i}-\frac{2\nabla_{\|}G_i}{3 n m_i}\label{eq_vpari}\\
%%%
%
% Electron temperature (metric)
%
 \frac{d T_e}{d t} = &\frac{4}{3}\frac{T_e}{eB}\left[\frac{7}{2}\hat{C}(T_e)+\frac{T_e}{n}\hat{C}(n)-
e\hat{C}(\phi)\right] \nonumber\\
&+\frac{2T_e}{3en}\left[0.71\nabla\cdot\left(j_\|\hat{\vec{b}}\right)-en\nabla\cdot\left(v_{\|e}\hat{\vec{b}} \right) \right]\nonumber\\
& -v_{\|e}\nabla_{\|}T_e+\chi_{\perp,e} \nabla_\perp^2 T_e + \chi_{\|,e} \nabla^2_\| T_e+ S_{T_e}\label{eq_tempe}
\end{align}In these equations, $df/dt=\partial_tf/\partial t+\left\{\phi,f\right\}/B$, we use the Poisson bracket $\left\{g,f\right\}=\vec{\hat{b}}\cdot\left(\nabla g\times \nabla f \right)$, and the curvature operator $\hat{C}(f)=(B/2)(\nabla\times(\hat{\vec{b}}/B))$. The unit magnetic field vector is $\hat{\vec{b}}=\vec{B}/B$, $j_\|=en(v_{\|i}-v_{\|e})$ is the parallel current, and $\sigma_\|$ is the Spitzer conductivity. The coordinate system is given by the poloidal length, radial, and toroidal angle coordinates $(y=\theta a,x,\varphi)$. $S_{T_e}$ and $S_n$ represent source terms used to inject density and temperature into the simulation domain. The numerical implementation of \ref{eq_dens}--\ref{eq_tempe}, including the definition of the gyroviscous terms $\sim G_{e,i}$ and other dissipative contributions, is described in detail in Ref.~\onlinecite{Halpern2016}. {(It has been checked that the artificial dissipation terms do not affect the simulation results.)} Sheath boundary conditions, modeling the interface between the SOL plasma and the vessel walls, are applied at the entrance of the magnetized pre-sheath where the ion drift approximation breaks down~\cite{Loizu2012}.% {\gbb Radial boundary conditions at the interface with the plasma core are $\partial_x \{n,\Omega,v_{\|i},v_{\|e},T_e\}= 0$ for all fields, except for $\phi$, which is set to a constant whose value does not affect flows or plasma gradients near the LCFS.}

%Here, we use the Poisson bracket $\left\{f,g\right\}=\vec{\hat{b}}\cdot\left(\nabla f\times \nabla g \right)$ and the curvature operator $\hat{C}(f)=(B/2)(\nabla\times(\hat{\vec{b}}/B))$, while $\hat{\vec{b}}=\vec{B}/B$ is the unit magnetic field and $j_\|=n(v_{\|i}-v_{\|e})$ is the parallel current. The radial, poloidal length, and toroidal coordinates are $(x,y,\varphi)$. $S_{T_e}$ and $S_n$ represent source terms used to inject plasma into the simulation domain. The equations are given in dimensionless form, with reference units $t\sim R/c_{s0}$, $\nabla_\perp \sim \rho_{s0}^{-1}$ ($\rho_{s0}=c_{s0}/\omega_{ci}$), $\nabla_\|\sim R^{-1}$, $B\sim B_0$, $n\sim n_{e0}$, $T_e\sim T_{e0}$, and $v_\|\sim c_{s0}$. The subscript 0 refers to the magnetic axis for $R_0$ and $B_0$, and to the last closed flux surface (LCFS) for $n$ and $T_e$. The dimensionless plasma parameters appearing in the equations are the normalized ion sound Larmor gyroradius $\rho_\star=\rho_{s0}/R$, and the dimensionless Spitzer resistivity $\nu=e n_{e0} R / (m_i c_{s0} \sigma_\|)$ ($\sigma_\|$ is the Spitzer conductivity). The exact definition of the gyroviscous terms $\sim G_{e,i}$ and other dissipative contributions are given in Ref.~\onlinecite{Halpern2016}. Sheath boundary conditions, modeling the interface between the SOL plasma and the vessel walls, are applied at the entrance of the magnetized pre-sheath where the ion drift approximation breaks down~\cite{Loizu2012}. The numerical implementation of Eqns.~\ref{eq_dens}--\ref{eq_tempe} is described in detail in Ref.~\onlinecite{Halpern2016}.

\begin{figure}
\includegraphics[width=0.45\textwidth]{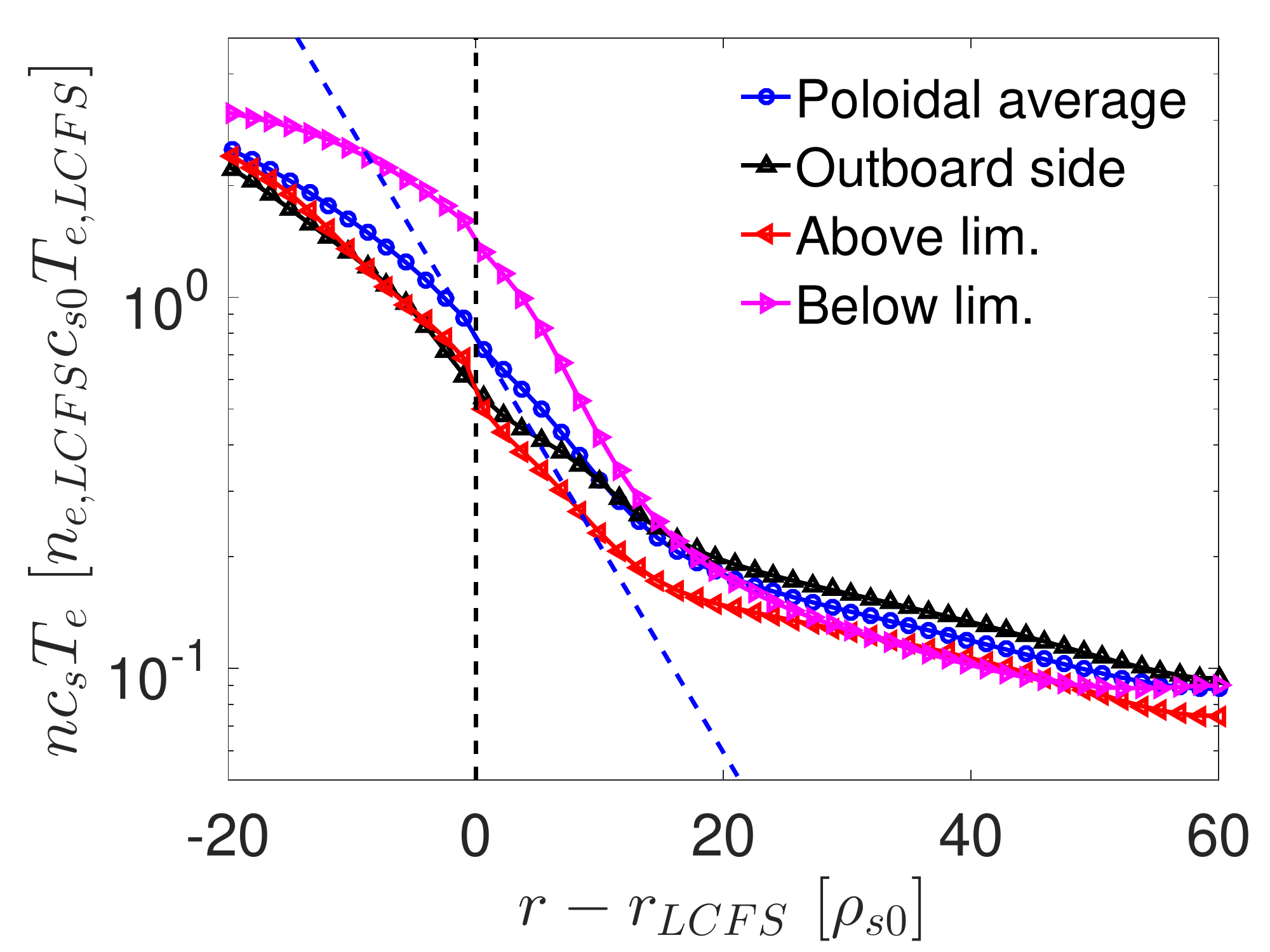}
\caption{Time averaged radial profiles of $nc_{s}T_e$, computed using (a) the entire poloidal cross-section (blue dotted line), (b) the equatorial outboard side of the device (black line with triangles), and (c) just above (red line with left-triangles) and (d) below the limiter (magenta line with right-triangles). Data obtained from the quasi steady-state phase of a simulation with $q=4$, $\rho_\star^{-1}=500$, $\nu=0.01$\label{fig_lq}.}
\end{figure}

Simulations are carried out within the parameter range $\rho_\star^{-1}=R/\rho_{s0}=250$--$1000$, $\nu=e^2 n_{e,LCFS} c_s/(m_i \sigma_\| R)=0.01,0.1,1$, $q=4$--$16$, $m_e/m_i=1/200$, with $a/R\approx 1/4$ ($q\approx (r/R)(B_\phi/B_\theta)$ is the magnetic safety factor, while $\rho_{s0}=c_{s0}/\omega_{ci}$, $c_{s0}=\sqrt{T_{e,LCFS}/m_i}$, and $\omega_{ci}=eB/m_i$). The simulation parameters $\rho_\star=2000$, $\nu=0.01$ roughly translate to the IWL SOL parameters of Alcator C-Mod ($R_0=0.67$m, $B_0=4$T, $T_{e,LCFS}=25$eV, $n_{e,LCFS}=10^{19}$m$^{-3}$). Using a simulation with $\rho_\star^{-1}=500$, $q=4$, and $\nu=0.01$, e.g. corresponding to C-Mod parameters but with $B_0=1$T, we illustrate the basic physics mechanisms giving rise to the narrow heat-flux feature. The simulation domain entails an annular volume representing the plasma edge and the SOL, where an infinitely thin wedge acts as a limiter on the high-field-side. Temperature and density are added within the plasma edge using poloidally uniform, radially Gaussian sources ($S_{T_e}$ and $S_n$) of radial width $5\rho_{s0}$ and placed at the inner boundary of the simulation domain. The plasma profiles steepen due to the action of the sources, driving turbulent modes that fill the SOL with plasma. Figure~\ref{fig_lq} shows steady-state, poloidally and toroidally averaged, radial profiles of $nc_sT_e$ showing a very clear break in slope about $20\rho_{s0}$ away from the LCFS ($nc_sT_e\sim q_\|$ near the limiter). The near SOL has $\lambda_q \approx 8\rho_{s0}$, which is equivalent to about 4mm in C-Mod ($B=4$T, $T_{e,LCFS}\approx 25$eV) and agrees with experimental measurements~\cite{Kocan2015}. From here onwards, we consider time, poloidally and toroidally averaged quantities (denoted with angled brackets $\left<\right>$) in order to highlight the main physical mechanisms at play.

%Figure~\ref{fig_profs} shows a poloidal cross-section of $\phi$ during the quasi-steady state phase of the simulations, where the plasma quantities are subject to power balance. Visibly, 

\begin{figure}
\includegraphics[width=0.45\textwidth]{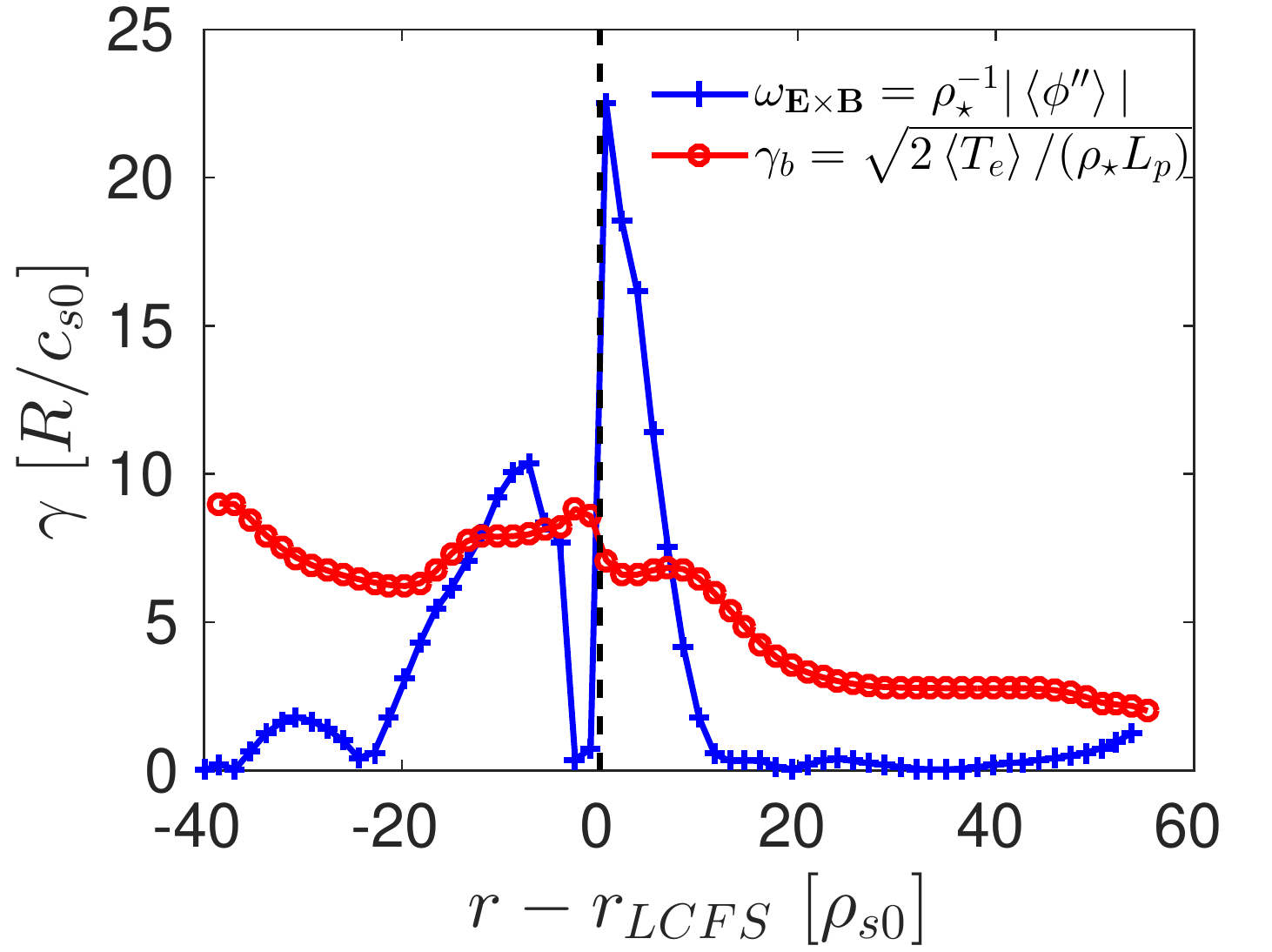}
\caption{Radial profiles of $\omega_{\vec{E}\times\vec{B}}=\rho_\star^{-1}\left|\left<\phi\right>^{\prime\prime} \right|$, and the ballooning growth rate, $\gamma_b= \sqrt{2\left<T_e\right>/(\rho_\star L_p)}$. Computed from a simulation with $q=4$, $\rho_\star^{-1}=500$, $\nu=0.01$\label{fig_wexb}.}
\end{figure}

The radial component of the steady-state electric field, $\left<E_x\right>=-\partial_x \left<\phi\right>$ has opposite signs in the SOL and in the plasma edge. In the SOL, the interaction between the plasma and the sheath gives $\left<\phi\right>\sim \Lambda\left<T_e\right>/e$ ($\Lambda\approx3$), \emph{i.e.} $\partial_x \left<\phi\right> >0$, while in the plasma edge $\partial_x \left<\phi\right> <0$. As a result, $<\phi>$ varies significantly around the LCFS, giving rise to a poloidal velocity shear layer in our simulations. In Fig.~\ref{fig_wexb}, the shearing rate $\omega_{\vec{E}\times\vec{B}}=\rho_\star^{-1}\left|\left<\phi\right>^{\prime\prime} \right|c_s/R$ is compared against the reference ballooning growth rate $\gamma_b=\sqrt{2\left<T_e\right>/(\rho_\star L_p)}c_s/R$ ($L_p=-d_x \ln \left< p\right>$). The shear layer effectively divides the edge of the plasma into 3 regions: (a) {\gbb the plasma edge, where $\gamma_b$ is comparable or larger than $\omega_{\vec{E}\times\vec{B}}$}, (b) the near-SOL, where drift and ballooning type modes are strongly linearly stable due to the velocity shear layer, and (c) the far SOL, where $\omega_{\vec{E}\times\vec{B}}$ is weak. The latter region was extensively described in our previous studies~\cite{Halpern2013,Halpern2014}.

%\begin{figure}
%%\includegraphics[width=0.45\textwidth]{ne_q4.png}
%\includegraphics[width=0.45\textwidth]{phi_q4.png}
%\caption{Poloidal cross-section of the electrostatic potential during the quasi steady-state phase of a simulation with $q=4$, $\rho_\star^{-1}=500$, $\nu=0.01$.\label{fig_profs}. The wedge on the inner side of the torus represents a toroidal limiter.}
%\end{figure}

\begin{figure}
\includegraphics[width=0.45\textwidth]{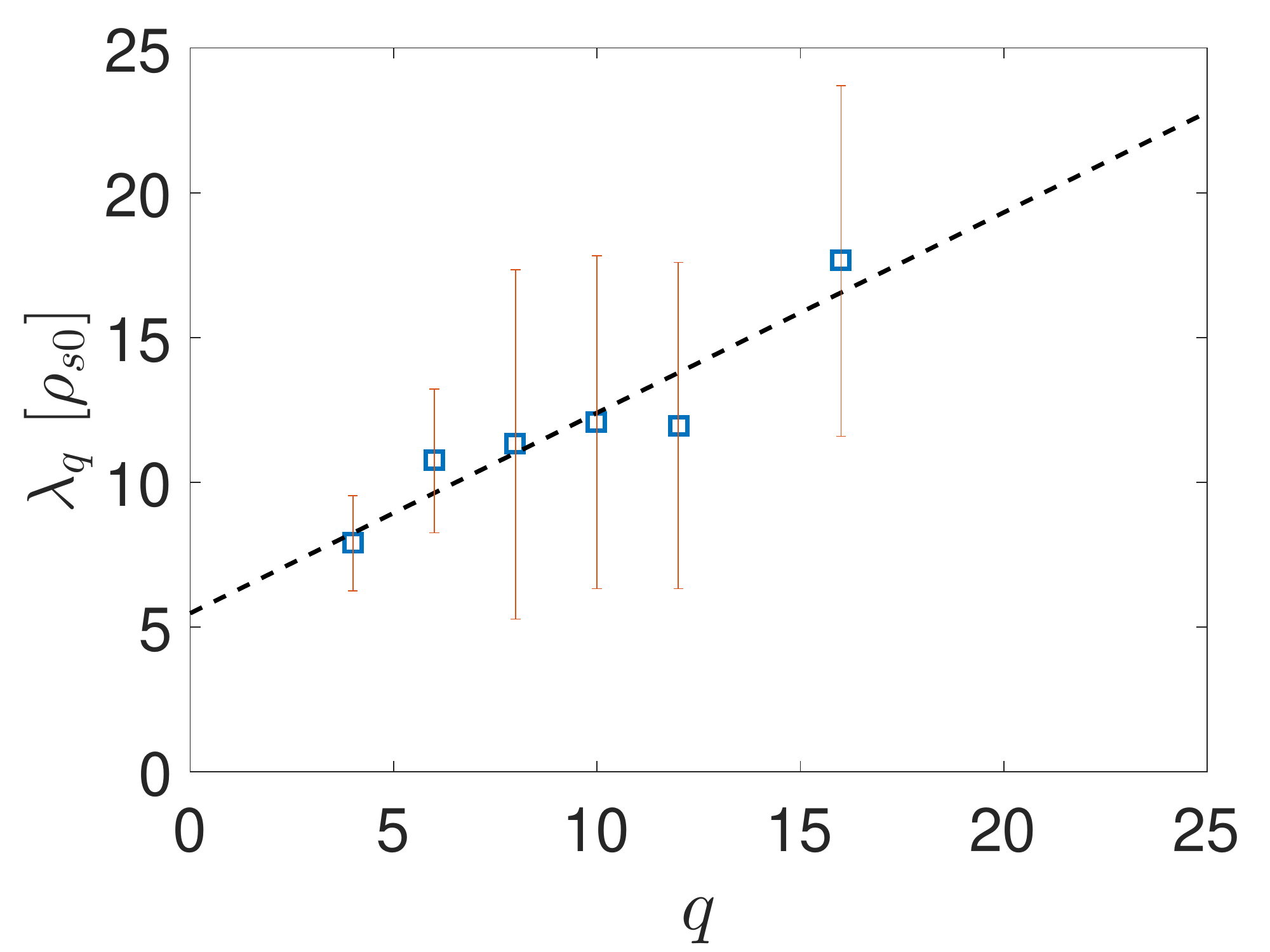}
\caption{
Simulated narrow-feature widths in simulations with $q=4$--$16$, $\rho_\star^{-1}=500$, $\nu=0.01$.\label{fig_Lq_vs_q}}
\end{figure}

We typically find $\Lambda \left<T_e\right> > \left<\phi\right>$ at the LCFS of our simulations, which is consistent with Langmuir probe measurements in the near-SOL of TCV and COMPASS~\cite{Nespoli2014,Dejarnac2015}. This phenomenon, in fact, suggests that parallel currents flowing out of the plasma play an important in the near-SOL, since by charge conservation $j_\perp / L_\perp \sim j_\| / L_\|$. This simple heuristic argument immediately relates the near-SOL width, which should be similar to $L_\perp$, to the safety factor $q\sim 1/B_\theta$. Indeed, a simulation scan over $q=4$--$16$, shown in Fig.~\ref{fig_Lq_vs_q}, {\gbb confirms that $\lambda_q/\rho_s \propto q$ at fixed $\nu=0.01$ and $\rho_\star^{-1}=500$}. The error bars give the root-mean-square deviation obtained from fitting $\left<nc_sT_e\right>$ over a time interval of 40$R_0/c_{s0}$.

%The strength of the effect is inversely proportional to the connection length $L_c=q$, pointing towards the $\lambda_q\sim 1/B_\theta$ scaling. 

Additional simulation scans have been carried out varying $\nu$ and $\rho_\star$ at fixed $q=4$. In the first case, the resistivity only has an effect when $\nu \sim 1$, in which case we observe weaker $j_\|$ near the limiter and an increased radial transport. Within the explored parameter range, we find little variation of $\lambda_q/\rho_s$ with $\rho_\star$, which suggests a weak dependence on the normalized plasma size. %Altogether, our simulations suggest $\lambda_q / \rho_s \propto q$, which is equivalent to Goldston's simplified heuristic drift formula~\cite{Goldston2012}.

\begin{figure}
\includegraphics[width=0.45\textwidth]{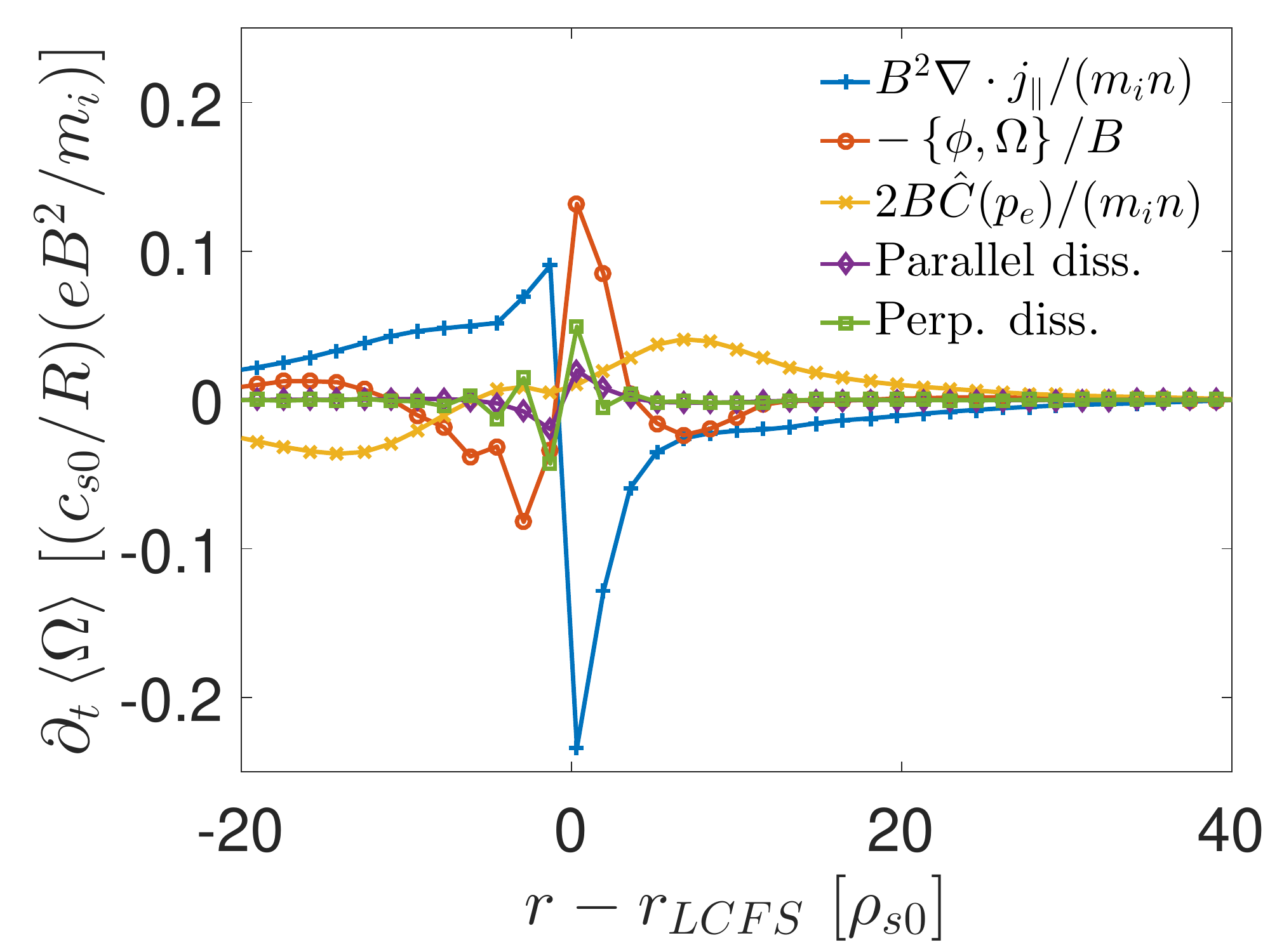}
\caption{Charge balance contributions from parallel currents (blue line with dots), vorticity convection (red line with crosses), curvature effects (yellow line with x's), parallel (purple line with diamonds) and perpendicular (green line with squares) dissipation terms. Computed from quasi steady-state phase of a simulation with $q=4$, $\rho_\star^{-1}=500$, $\nu=0.01$.\label{fig_omega_bal}}
\end{figure}

To gain further insight on the role of the outflowing currents, we concentrate on the charge balance in the system, Eq.~\ref{eq_vort}. This is illustrated in Fig.~\ref{fig_omega_bal}, where we have separated the contributions of all the terms in the vorticity equation (including numerical dissipative terms), as radial profiles. We observe that the parallel current contribution, $\left<B^2 \nabla\cdot j_{\|}/(nm_i)\right>$, strongly affects the charge balance in the near-SOL. The parallel currents are mostly compensated through a polarization contribution $\sim\left<\left\{\phi,\Omega\right\}/B \right>$, while other terms play a minor role. The curvature term $2\left<B\hat{C}(p_e)/(nm_i)\right>$ plays an important role in the far-SOL, consistent with blob filament motion~\cite{KRASHENINNIKOV2008}. On the other hand, the radial dissipative terms become noticeable near the LCFS due to the steep gradients of the radial $\left<\Omega\right>$ profile -- {\gbb it has been tested that decreasing the radial diffusion steepens the profile by about 1$\rho_{s0}$, which is within the 95\% confidence interval of the $\lambda_q$ fit.}

%\section{Analytical model}\label{sec_analytical_model}

%It is noted, first, that {\gbb the linear growth rate of ballooning modes} is much smaller than $\omega_{\vec{E}\times\vec{B}}$. Thus, the typical assumption that a balance between linear growth and non-linear plasma convection leads to saturation (e.g. wave-breaking or gradient removal~\cite{Horton1999}) cannot be applied in the near-SOL. Furthermore, since the strong flow shear inhibits linear instability within the near-SOL, the perturbed electrostatic potential cannot be related to pressure perturbations through a linearized continuity equation.

%We contest, in particular, the idea that the perturbed electrostatic potential can be related to the density perturbation through a continuity or particle balance equation. This saturation model has been used extensively and with much success in the plasma physics community~\cite{Horton1999}, and more recently applied to SOL transport~\cite{Ricci2013}.

We now propose a reduced model predicting $\lambda_q$, based upon a balance between the $j_\|$ and $j_\bot$ contributions. Our objective is to obtain the transport levels within the narrow heat-flux feature. The perturbed electrostatic potential is determined through the vorticity balance, allowing us to evaluate the near-SOL $\vec{E}\times\vec{B}$ velocity. Consider a steady-state equation balancing parallel and polarization current terms at the LCFS. Integrating along the field line, and neglecting parallel mode anisotropy, we obtain
\begin{align}
\frac{1}{B^2}\left\{\phi,\Omega\right\}&= \frac{c_s \omega_{ci} }{L_\|}\exp\left(\frac{e\delta\phi_{fl}}{T_e} \right),
\end{align}where we have used the Gauss' theorem and simplified the sheath current $j_{sh}= enc_s(1-\exp(\Lambda-e\phi/T_e))\approx enc_s \exp( e\delta\phi_{fl}/T_e)$ ($\delta\phi_{fl}=\Lambda T_e/e -\phi$). The simulation results indicate that the polarization current contribution is dominated by a radially sheared convection of vorticity. Taking a poloidal average, we recover the expression
%Taking $B\approx B_0=1$ and $n=n_0=1$ at the LCFS, the field-line averaged charge balance equation simplifies to
\begin{align}
\left<\frac{1}{B^2} \frac{\partial}{\partial x} \left( \tilde{\Omega} \frac{\partial \tilde{\phi}}{\partial y}\right)\right>\approx \left< \frac{c_s\omega_{ci}}{L_\|}\exp\left(\frac{e\delta\phi_{fl}}{T_e}\right) \right>,\label{eq_qbal}
\end{align}with the tildes indicating perturbed quantities. This step points out that it is the radial shear of the turbulent motion that allows diverging parallel currents to arise. The currents flowing into the sheath, in turn, allow the potential to decouple from the temperature profile. The interaction with the closed magnetic field line region, where the electric field has the opposite sign than in the SOL, leads thereafter to the radially sheared electric field characteristic of the narrow heat-flux feature.

\begin{figure}
\includegraphics[width=0.45\textwidth]{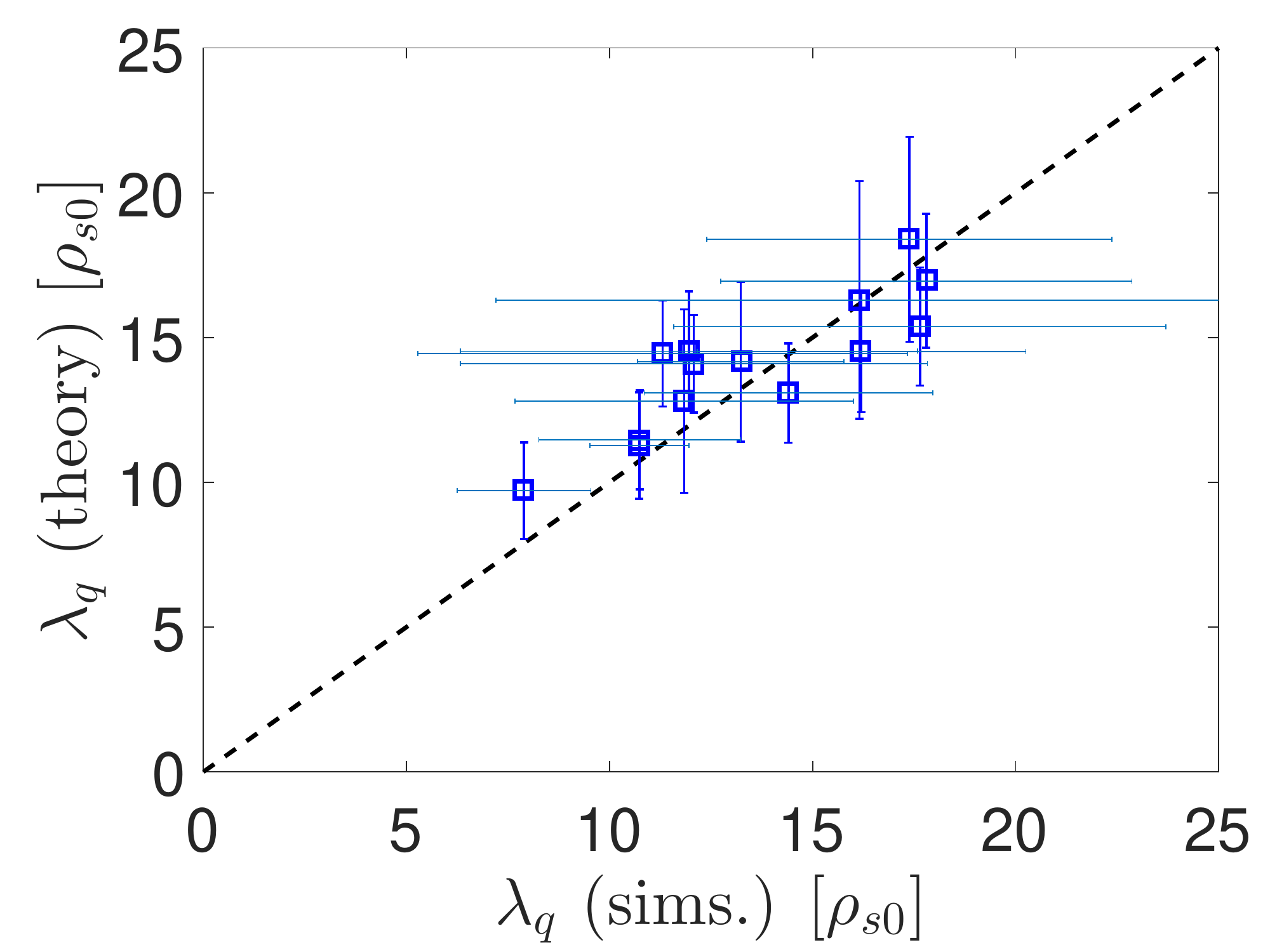}
\caption{Equation~\ref{eq_lq} is compared against non-linear simulation results.\label{fig_eq9}}
\end{figure}

Next, we estimate $\tilde{\Omega}=-k_\perp^2\tilde{\phi}$, and $\partial_x \sim k_x$ and $\partial_y \sim k_y$, which leads to the radial $\vec{E}\times\vec{B}$ velocity of turbulent structures propagating across the narrow feature
\begin{align}
\left<\tilde{v}_{\vec{E}\times\vec{B},x}\right>^2 \approx \left< \frac{c_s\omega_{ci}}{L_\|}\frac{k_y}{k_x k_\bot^2} \exp\left(\frac{e\delta\phi_{fl}}{T_e}\right) \right>.\label{eq_qvel}
\end{align}The turbulent flux follows immediately from the estimate $\left<\Gamma_\perp\right> \approx \left< \tilde{p}\tilde{v}_{\vec{E}\times\vec{B},x}\right>$. The amplitude of the fluctuations traversing the narrow feature from the edge is estimated as $\left<\tilde{p}\right> \sim \left<p\right>/(k_x \lambda_q)$~\cite{Horton1999,Ricci2013}. Then, the near-SOL width can be obtained by balancing $\nabla\cdot\left<\Gamma_\perp\right>$ against the sheath contribution $\nabla_\|\cdot \left< \Gamma_\| \right> \approx \left< p c_s \exp(e\delta\phi_{fl}/T_e)\right> / L_\|$. The assumption of parallel convection rather than conduction is justified in the case of weak poloidal plasma gradients, which was an assumption of our analysis. The result is
\begin{equation}
\lambda_q = \left<\frac{k_y}{k_x^3 k_\bot^2}\frac{L_\|}{\rho_s}\exp\left(\frac{-e\delta\phi_{fl}}{T_e} \right)\right>^{1/4}\approx \frac{k_x^{-1}}{2}\left(\frac{q}{\rho_\star}\right)^{1/4}.\label{eq_lq}
%L_q = \left<\frac{k_y}{k_x^3 k_\bot^2}\frac{L_\|}{\rho_\star}\frac{1}{c_s\rho_s^2}\frac{\delta\phi_{fl}}{B} \right>^{1/4}\approx k_x^{-1}\left(\frac{L_\|}{\rho_\star}\right)^{1/4}\label{eq_lq}
%L_q = \left( \frac{k_y q}{k_x^3 k_\perp^2\rho_\star}\right)^{1/4} \approx k_x^{-1} (q/\rho_\star)^{1/4}\label{eq_lq}.
\end{equation}In the last expression, we replaced $L_\|=qR$ and we assumed that eddys have comparable radial and poloidal wavenumbers, \emph{i.e.} $k_x\sim k_y \sim k_\perp$ around the LCFS. {\gbb The near SOL wavenumber is consistent with simulation results, and with gas-puff imaging of SOL turbulence~\cite{Zweben2007}. As the modes traverse into the far SOL, $k_x$ decreases while $k_y$ remains about constant.} We also approximate $\exp(-e\delta\phi_{fl}/T_e)^{1/4}\approx 1/2$, based on the LCFS values consistently found throughout our simulation scan. The weak dependence obtained with respect to the plasma parameters can explain, in part, why it is difficult to vary the narrow feature width in experiments -- the plasma parameters appear only indirectly, and through the radial correlation length $L_{rad}=\pi/k_x$. Equation~\ref{eq_lq} is the principal result of the model, and the simpler expression involving $k_x^{-1}$ is evaluated using the radial eddy correlation length and compared against non-linear simulation results in Fig~\ref{fig_eq9}.

In conclusion, we propose that a narrow layer of radially-sheared poloidal flows, occurring within the near-SOL, is responsible for the steep plasma gradients recently measured in the IWL tokamak experiments. Non-linear, flux-driven turbulent simulations demonstrate the spontaneous formation of $\vec{E}\times\vec{B}$ shearing rates significantly surpassing the expected linear growth rate of the turbulent modes. Simulation results suggest that $\lambda_q/\rho_s$ increases with $q\sim I_p^{-1}$, with weaker variation of $\lambda_q$ with respect to $\nu$ or $\rho_\star$. The analysis of the simulations leads us to conclude that the turbulent saturation level can be determined by balancing the polarization currents driven by the turbulence against the parallel currents observed at the limiter. Analytical estimates lead to a gradient length of the order of the turbulent correlation length. {\gbb The proposed transport model would suggest that a $\lambda_q\sim q\sim I_p^{-1}$ scaling (e.g. as in the Drift Heuristic Model~\cite{Goldston2012}) can originate from the turbulent wavenumber. Inertial ballooning modes (IBM) are the most linearly unstable modes in the parameter regime $q=4$, $\rho_\star^{-1}=500$, $\nu\approx0.01$ and with steep plasma gradients~\cite{Mosetto2012}. For instance, the wavenumber $k_{IBM}\rho_s \propto q^{-1}\gamma_b^{-1}$ together with Eq.~\ref{eq_lq} yield $\lambda_{q,IBM}/\rho_s \sim q^{5/6} \rho_{\star}^{-1/2} \nu^{0}$.}
%. This in turn would lead to $\lambda_{q,IBM}/\rho_s \sim q^{5/6} \rho_{\star}^{-1/2} \nu^{0}$}% For example, invoking typical wavenumbers for resistive (RBM) or inertial (IBM) ballooning modes ($k_{RBM}\rho_s \propto q^{-1} \nu^{-1/2} \gamma_b^{-1/2}$ or $k_{IBM}\rho_s \propto q^{-1}\gamma_b^{-1}$) we find to $\lambda_{q,RBM}/\rho_s \sim q \rho_{\star}^{-2/5} \nu^{-2/5}$ and $\lambda_{q,IBM}/\rho_s \sim q^{5/6} \rho_{\star}^{-1/2} \nu^{0}$. It is noted, however, that the modes around the LCFS have $\tilde{n}\sim\tilde{\phi}$, \emph{i.e.} they are probably drift-type modes rather than ballooning.}

As a final remark, we highlight that our results lead to several testable predictions: (a) the turbulent intensity allows the outflow of parallel currents at the limiter, (b) the strength of the currents can be related to $k_x$, and (c) $\lambda_q$ decreases with $q^{-1}\sim B_\theta$. Some of these features, such as the currents at the contact points, have been observed before in several devices. Dedicated experimental campaigns at C-Mod, DIII-D, and TCV will be used with the objective of validating the physical insights here presented.

\acknowledgements
Part of the simulations presented herein were carried out using the HELIOS supercomputer system at the Computational Simulation Centre of International Fusion Energy Research Centre (IFERC-CSC), Aomori, Japan, under the Broader Approach collaboration between Euratom and Japan, implemented by Fusion for Energy and JAEA. This work has been carried out within the framework of the EUROfusion Consortium and has received funding from the European Union's Horizon 2020 research and innovation program under grant agreement number 633053, and from the Swiss National Science Foundation. The views and opinions expressed herein do not necessarily reflect those of the European Commission.
%; and part were carried out at the Swiss National Supercomputing Centre (CSCS) under Project ID s549. This research was supported by the Swiss National Science Foundation.

%\bibliographystyle{aipnum4-1}
\bibliography{SOL_EM}
%\bibliography{/misc/sandbox/Research/Manuscripts/Bibliography/SOL_EM.bib}
\end{document}